\documentclass[aps,prb,superscriptaddress,twocolumn,floatfix]{revtex4-2}
\usepackage[abs]{overpic}
\usepackage{graphicx,graphics}
\usepackage{dcolumn}
\usepackage{amsmath,amssymb,amsfonts}
\usepackage{wasysym}
\usepackage{latexsym,verbatim}
\usepackage{bm}
\usepackage{color}
\usepackage{ulem}
\usepackage[framemethod=default]{mdframed}
\usepackage[breaklinks=true,colorlinks,citecolor=blue,linkcolor=blue,urlcolor=blue]{hyperref}
\usepackage[makeroom]{cancel}
\usepackage{fancybox}

\newcommand{\be}{\begin{eqnarray}}
\newcommand{\ee}{\end{eqnarray}}

\begin{document}

\preprint{APS/123-QED}

\title{Emergent Hyper-Magic Manifold in Twisted Kitaev Bilayers}

\author{Samuel Haskell}
\email{samuel.haskell@manchester.ac.uk}
\author{Alessandro Principi}
\email{alessandro.principi@manchester.ac.uk}

\affiliation{
 Department of Physics and Astronomy, University of Manchester, Manchester, M13 9PL, UK
}

\begin{abstract}
Kitaev quantum spin liquids have been the focus of intense research effort thanks to the discovery of various materials (e.g., RuCl$_3$) that approximate their intriguing physics. In this paper we construct a mean-field approximation for a moir\`e superlattice emerging in twisted Kitaev bilayers in terms of solutions of commensurate bilayers. We show that the band structure of deconfined spinons, defined on the mini-Brillouin zone of the superlattice, is greatly modified. The system exhibits a hyper-magic manifold: a series of nearly perfectly-flat bands appear at energies above the lowest gap, exhibiting a very large (spinon) density of states that could potentially be probed experimentally.
Intriguingly, flat-band eigenstates exhibit a localization akin to wavefunctions of Kagome lattices.
\end{abstract}

\maketitle

{\it Introduction}---Kitaev quantum spin liquids (QSLs) are forms of topological order~\cite{Wen2002} that support fermionic and Majorana spinon excitations~\cite{Kitaev2006}. They have recently attracted renewed interest upon discovery that RuCl$_3$, a transition-metal halide~\cite{Huang2017,Wang2018,Sivadas2018,Gong2017,Klein2018,Yin2018,Huang2018} Van-der-Waals material~\cite{Geim_Nature_2013,Novoselov_Science_2016}, supports Kitaev QSL phases~\cite{Kitaev2006,PhysRevLett.109.085302,PhysRevLett.102.017205,PhysRevLett.109.085303,PhysRevB.90.041112,annurev-conmatphys-033117-053934,Kitaev2006,PhysRevLett.107.077201,Balents_nature_2010,Wen2002,PhysRevB.90.041112,Zhou2018,Ran2017,Banerjee2017,Banerjee2016,Zhou2018a,Sandilands2015,PhysRevLett.123.237201,Iyikanat2018,PhysRevB.90.041112,Sandilands2015,PhysRevLett.123.237201,Iyikanat2018,PhysRevB.90.041112,PhysRevB.95.180411,PhysRevB.102.155134,PhysRevB.103.184407,PhysRevResearch.2.013014,Gordon2019,Zhang2022,Lee2020,Janssen2019,Gordon2021,Li2021}. Neutron-scattering~\cite{Banerjee2017,Banerjee2016}, Raman~\cite{Zhou2018a}, and thermal Hall~\cite{Kasahara2018,PhysRevLett.120.217205,PhysRevB.99.085136,Yokoi2021,Bruin2022,PhysRevB.102.220404} measurements all support this picture. In this paper we are concerned with the broad phenomenology emerging from the interplay between quantum ordering and superlattices. In particular, long-periodicity moir\`e superlattices can greatly modify the properties of the whole heterostructure~\cite{PhysRevLett.122.106405,Kerelsky2019,Yankowitz2019,Zheng2020,Haddadi2020,Kim2017,Carr2020,McGilly2020,Balents2020,Cao_nature_2018_2,Cao_nature_2018,Park2021,Pierce2021,Xie2021,Hesp2021,Cao2020} (see also Refs.~\cite{He2021,Andrei2021} and references therein). The scope of this Letter is therefore to propose a methodology bridging the microscopic scales of short-range interactions (which lead to the emergence of QSLs and exotic quasiparticles) with those of superlattices encompassing hundreds of unit cells. We focus on the exactly-solvable Kitaev model~\cite{Kitaev2006}: a hexagonal quantum spin-$1/2$ system with bond dependent Ising-like interactions. This model provides a qualitative understanding of the low-energy physics~\cite{PhysRevLett.102.017205} of RuCl$_3$.
\begin{figure}[t!]
    \centering
    \includegraphics[width=\linewidth]{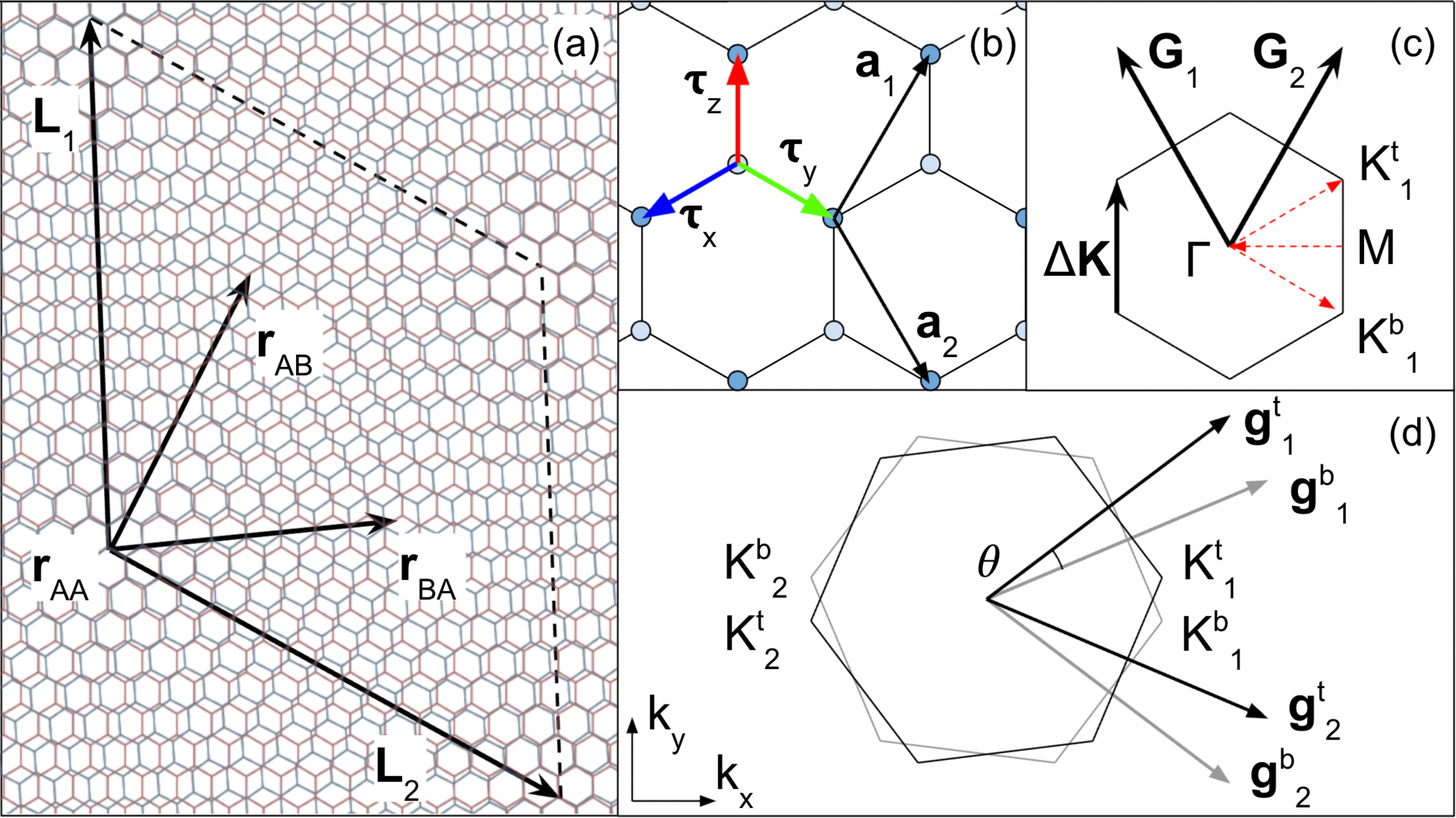}
    \caption{
Panel (a) Moir\`e pattern generated from two hexagonal lattices, with relative twist $\theta = 4^{\circ}$. The supercell, spanned by vectors $\bm{L}_{1, 2}$ is outlined. Approximately commensurate AA, AB and BA stacking configurations occur at $\bm{r}_{\rm AA}$, $\bm{r}_{\rm AB}$ and $\bm{r}_{\rm BA}$ respectively. 
Panel (b) Basis vectors $\bm{a}_{1,2}$  for the individual hexagonal layers. $\bm{\tau}_{x}$, $\bm{\tau}_{y}$, and $\bm{\tau}_{z}$ are the nearest-neighbor vectors. Colored arrows denote different bond types, along which different spin components are coupled.
Panel (c) MBZ generated by vectors $\bm{G}_{1,2}$. $\Delta \bm{K}$ connects points $K^t_1$ and $K^b_1$, {\it i.e.} the Dirac points of each layer. 
Panel (d) Brillouin zones for the top (black) and bottom (grey) layers, twisted and superimposed. Reciprocal lattice vectors $\bm{g}_{1,2}$ and inequivalent Dirac points $\bm{K}_{1, 2}$ for each layer are indicated.
}
\label{fig:one}
\end{figure}
Commensurate Kitaev bilayers, obtained by coupling two such hexagonal lattices, have been addressed in Refs.~\cite{Seifert2018,Okamoto_prb_2013,Tomishige2018,Tomishige_prb_2019, PhysRevB.104.115123}. In particular, Ref.~\cite{Seifert2018} details a mean-field solution of bilayers with perfect atom-on-atom stacking. There, it is shown that both the intralayer and interlayer hopping amplitudes of Majorana excitations depend on the interlayer spin coupling $J$. For small $J$, the two layers are {\it exactly} decoupled, {\it i.e.} spinons {\it cannot} tunnel between the two. On the contrary as $J$ increases (within the mean-field approximation, to about $\sim 0.44$ times the intralayer Kitaev coupling $K$), interlayer hopping turns on and grows with $J$. Notably, when this happens the {\it intralayer} hopping decreases, {\it i.e.} spinons are less mobile in each layer, and vanishes for $J \simeq 0.58K$. For larger $J$, the two layers are strongly coupled and spinons {\it cannot} propagate. We find that in AB bilayers, where only half of the atoms of one layer sit on top of atoms of the other layer, the boundaries between the three phases shift to approximately twice the corresponding values of $J/K$ reported for AA bilayers. Mean-field results have been extended with a variety of advanced techniques (see, e.g., Refs.~\cite{Koga_JSNM_2019,PhysRevResearch.3.013160,PhysRevB.102.214422,PhysRevB.104.115123,PhysRevB.99.174424,PhysRevB.97.094403}). However, since our aim is to present a methodology to capture {\it qualitative} changes of twisted-stack properties, and to keep the problem tractable, we will employ mean-field results~\cite{Seifert2018} for commensurate bilayers. Other Kitaev heterostructures have also been studied with techniques similar to those mentioned above~\cite{You2012, Choi2018, Seifert2018_2, Halasz2014}.

We construct a mean-field approximation by interpolating the results for commensurate AA and AB stacking with the periodicity of the moir\`e superlattice. For certain ranges of parameters, spinons can tunnel between layers in AA regions but not in AB regions. In turn, their intralayer hopping is suppressed in AA regions, up to the point where it can vanish, but remains unaffected in AB regions. This reflects on the band structure of excitations: a single band around zero energy becomes separated from the other ones. Furthermore, the energy spectrum exhibits a hyper-magic manifold~\cite{Scheer_arxiv_2022}: some higher-energy bands can become extremely flat, even before the flattening of the low-energy one occurs. These bands exhibit an extremely high density of states, which could be probed in 2D-to-2D tunneling experiments~\cite{carrega_prb_2020}. Their eigenstates exhibit localization patterns resembling Kagome lattices, which could be observed in spin-resolved scanning tunneling spectroscopy. Our results are distinct from those of previous studies~\cite{May-Mann2020}, which focused on the regime of strong interlayer coupling. Furthermore, our model provides a physical realization of the hyper-magic manifold described in Ref.~\cite{Scheer_arxiv_2022}.

{\it Twisted Bilayer Kitaev Model}---A moir\`e pattern emerges when two copies of an identical lattice are superimposed with some relative twist angle $\theta$ between them, resulting in a periodic repetition of different commensurate stacking configurations. At small twist angles the pattern is approximately periodic, with a well-defined moir\`e supercell~\cite{Bistrizer2011}. Fig.~\ref{fig:one}(a) shows the canonical example of two hexagonal lattices of identical atoms and identical bonds, with periodic repetition of AA, AB and BA stacking configurations. In the following we consider a moir\`e heterostructure consisting of two Kitaev monolayers, with a relative twist between them.

The monolayer Kitaev model is defined on a hexagonal lattice, whose unit cell is spanned by a pair of vectors $\bm{a}_{1,2} = (1/2, \pm \sqrt{3}/2)$. The vectors $\bm{\tau}_{z} = (\bm{a}_1 - \bm{a}_2)/3$, $\bm{\tau}_{y} = (2\bm{a}_2 + \bm{a}_1)/3$, $\bm{\tau}_{x} = -(\bm{a}_2 + 2\bm{a}_1)/3$ connect elements of the A sublattice to their nearest neighbours on the B sublattice [see Fig.~\ref{fig:one}(b)]. The reciprocal lattice is spanned by the vectors $\bm{g}_{1,2} = b(\sqrt{3}/2, \pm 1/2)$, where $b = 4\pi/\sqrt{3}$. These define the maximally-symmetric hexagonal Brillouin zone (BZ) with corners at inequivalent points $\bm{K}_{1, 2} = \pm(\bm{g}_1 + \bm{g}_2)/3$.  Spin-$1/2$ quantum variables are located at each site, coupled via bond-dependent Ising interactions~\cite{Kitaev2006}. The $x$, $y$ and $z$ components of the spins couple along the blue, green and red directions respectively, as shown in Fig.~\ref{fig:one}(b). The solution of the Kitaev model passes through the {\it fractionalization} of each spin variable in terms of four Majorana particles~\cite{Kitaev2006}. Of these, three are paired between neighboring sites to form a gauge field upon which the fourth (deconfined) one propagates (more details are given below). Since the gauge potential commutes with the Hamiltonian and with itself on different bonds, the model reduces to an exactly-solvable quadratic Hamiltonian for deconfined particles~\cite{Kitaev2006}. The energy dispersion is akin to graphene's, {\it i.e.} it is gapless and exhibits Dirac cones at the points $\bm{K}_{1, 2}$.

The twisted heterostructure is generated from a commensurate bilayer structure by applying rigid rotations $\mathcal{R}(\pm \theta/2)$ to the top and bottom layers respectively. At zero twist angle, bonds of equal type [{\it i.e.} $x$, $y$, $z$ or red, green, blue in Fig.~\ref{fig:one}(b)] are chosen to be exactly aligned on top of each other. The resulting moir\`e pattern has an associated reduced moir\`e Brillouin zone (MBZ) in momentum space [Fig.~\ref{fig:one}(c)], and reciprocal lattice spanned by vectors $\bm{G}_{1,2} \equiv \bm{g}^{t}_{1,2} - \bm{g}^{b}_{1,2}$, where $\bm{g}^{t, b}_{\alpha} \equiv \mathcal{R}(\pm \theta/2)\bm{g}_{\alpha}$ are the rotated vectors of the top/bottom layers [Fig.~\ref{fig:one}(d)]. The real-space superlattice is thus characterized by vectors $\bm{L}_{1,2}$ [Fig.~\ref{fig:one}(a)] satisfying $\bm{L}_i \cdot \bm{G}_j = 2\pi \delta_{ij}$. For small angles one can approximate each region of the heterostructure by the corresponding commensurate stacking. 

Commensurate bilayers are uniquely characterised by the in-plane displacement $\bm{d}$ of the top layer relative to the bottom one. AB stacking occurs where $\bm{d} = \bm{\tau}_{z}$, such that the A sublattice in the top layer is aligned with the B sublattice in the bottom layer. Similarly, BA stacking occurs where $\bm{d} = -\bm{\tau}_{z}$, and AA stacking occurs where $\bm{d} = \bm{0}$. The positions in the supercell where the local structure approximates the AA, AB, and BA stackings are $\bm{r}_{\rm AA} = \bm{0}$, $\bm{r}_{\rm AB} = (2\bm{L}_{1} + \bm{L}_{2})/3$, and $\bm{r}_{\rm BA} = (\bm{L}_{1} + 2\bm{L}_{2})/3$ respectively [Fig.~\ref{fig:one}(a)]. Commensurate bilayer Kitaev models have previously been studied using mean-field theory. Mean-field solutions modify the hopping amplitudes of deconfined Majorana particles. Given that the stacking arrangement influences the resulting mean-field solution, one  expects the value of tight-binding parameters in a twisted bilayer to vary spatially with the superlattice. In twisted bilayer graphene~\cite{He2021,Andrei2021}, these spatial variations enter only via the interlayer hopping amplitudes, while the intralayer dynamics are unchanged. On the other hand, in the Kitaev model the interlayer and intralayer dynamics cannot effectively be disentangled, leading to novel twist-dependent intralayer terms.

We propose a model of the twisted Kitaev bilayer analogous to the continuum model of twisted bilayer graphene~\cite{Bistrizer2011}. The full Hamiltonian is a sum of two Kitaev monolayers $H^{t, b}$ coupled by a Heisenberg-like interlayer interaction $H^{\perp}$ of strength $J$, such that $H = H^{t} + H^{b} + H^{\perp}$. Each layer $l=t,b$ is described by an isotropic hexagonal spin-$\frac{1}{2}$ Kitaev model
\begin{equation}
    \label{eq:Kitaev}
    H^{l}_{K} = - \frac{1}{2} \sum_{\alpha} \sum_{i} K_{\alpha} S^{\alpha}_{\bm{r}_{i}, l, A} S^{\alpha}_{\bm{r}_{i}+\bm{\tau}^{l}_{\alpha}, l, B},
\end{equation}
where $i$ labels the unit cell, and $K_{\alpha}$ is a bond-dependent Kitaev coupling. Vectors with superscript $l$ are rotated according to their layer as $\bm{v}^{t, b} \equiv \mathcal{R}(\pm\theta/2)\bm{v}$, while $\bm{r}_i$ is a coordinate defined in a global reference frame. Each spin may be represented by a set of four Majorana fermions $\{c, b^{x}, b^{y}, b^{z}\}$ such that $S^{\alpha} \equiv ib^{\alpha} c$. Moving to the Majorana fermion representation and performing a mean-field decoupling \cite{Seifert2018, You2012}, one then obtains
\begin{equation} \label{eq:Ham_MF}
\begin{split}
    H^{l}_{K} & = \frac{i}{2} \sum_{\alpha} \sum_{i} K_{\alpha} \bigl( u^{l}_{\alpha}(\bm{r}_{i}) c_{\bm{r}_{i}, l, A} c_{\bm{r}_{i}+\bm{\tau}^{l}_{\alpha}, l, B} \\ 
    & + u^{l}_{0, \alpha}(\bm{r}_{i}) b^{\alpha}_{\bm{r}_{i}, l, A} b^{\alpha}_{\bm{r}_{i}+\bm{\tau}^{l}_{\alpha}, l, B} \bigr),
\end{split}
\end{equation}
where the mean-field components are defined as $u^{l}_{\alpha}(\bm{r}_{i}) \equiv \langle i b^{\alpha}_{\bm{r}_{i}, l, A} b^{\alpha}_{\bm{r}_{i}+\bm{\tau}^{l}_{\alpha}, l, B} \rangle$ and $u^{l}_{0, \alpha}(\bm{r}_i) \equiv \langle i c_{\bm{r}_{i}, l, A} c_{\bm{r}_{i}+\bm{\tau}^{l}_{\alpha}, l, B} \rangle$ respectively. One may also exact a similar procedure on the interlayer coupling term $H^{\perp}$, yielding mean-field components $w_{\alpha}(\bm{r}_{i})$ and $w_{0, \alpha}(\bm{r}_{i})$. 
For structures studied here, we give details in the supplementary material~\footnote{See Supplemental Online Material.}.

\begin{figure}
    \centering
\begin{overpic}[width=0.9\columnwidth]{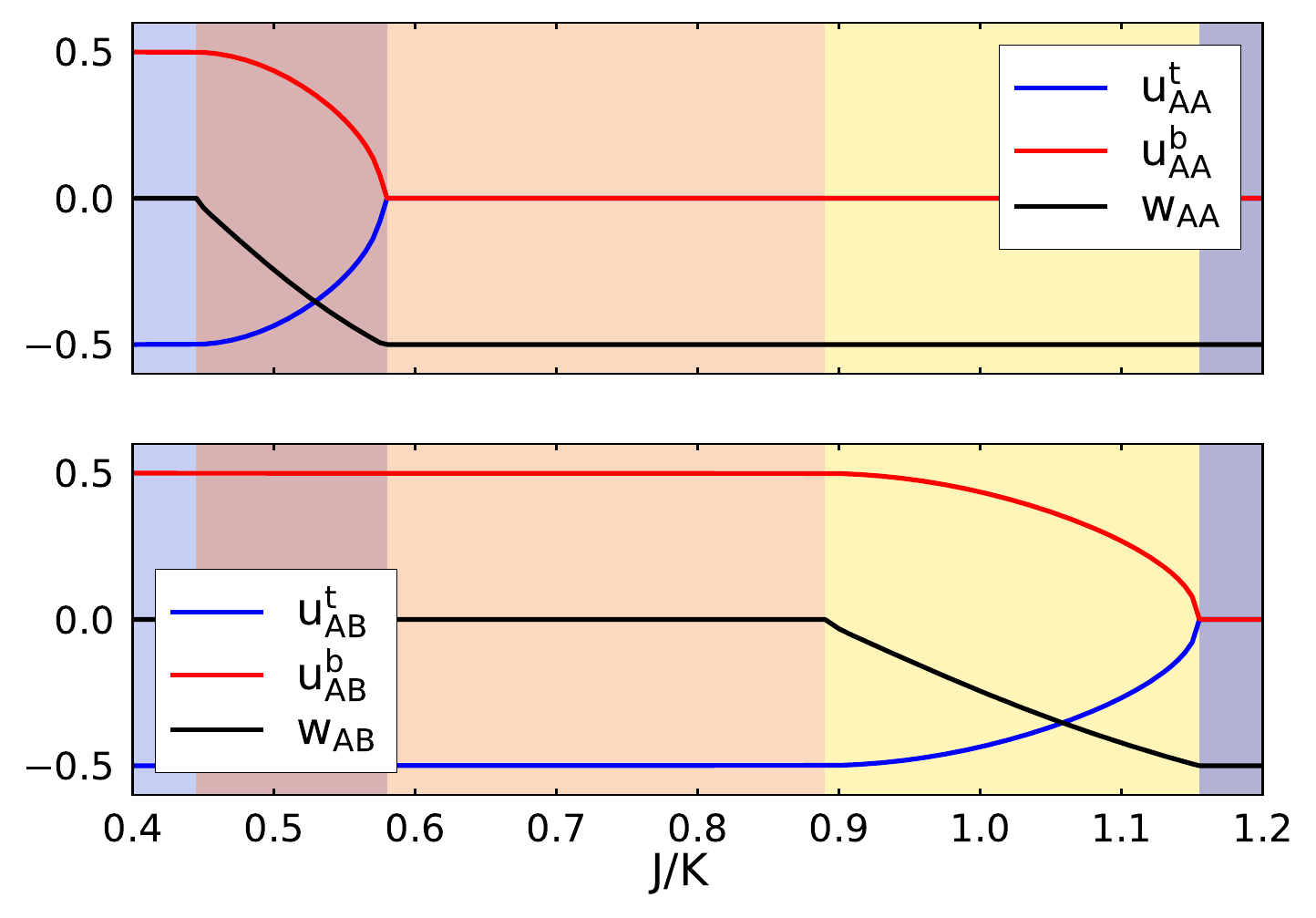}
\put(230,140){(a)}
\end{overpic}
\begin{overpic}[width=\columnwidth]{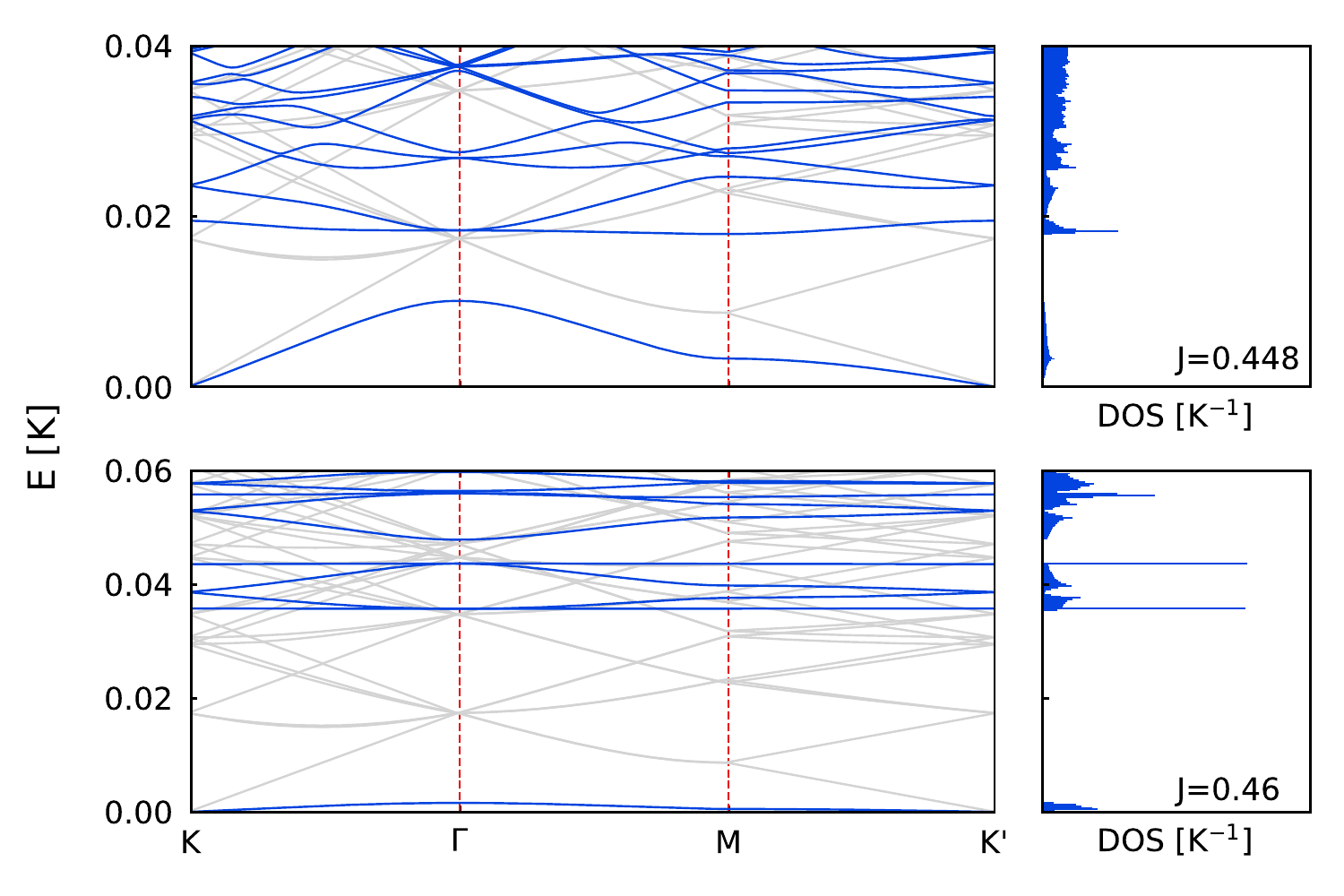}
\put(240,161){(b)}
\put(240,83){(c)}
\end{overpic}
\vspace{-8mm}
    \caption{
Panel (a) Mean-field solutions for AA (top) and AB (bottom) stacked bilayers as a function of $J/K$ in the isotropic limit. 
Panel (b) In blue, the spectrum of itinerant Majorana fermions for $J/K = 0.448$. Twist angle $\theta = 1.1^{\circ}$. The grey background represents the spectrum of the decoupled twisted bilayer ($J/K = 0$). 
Panel (c) Same as Panel (b), but for $J/K = 0.46$. $K$ and $K'$ are $K^t_1$ and $K^b_1$ of Fig.~\ref{fig:one}(c) respectively.}
    \label{fig:two}
\end{figure}

In Eq.~(\ref{eq:Ham_MF}) we have effectively separated the full quartic Hamiltonian into four separate (quadratic) tight-binding problems, one for each Majorana species. The effective Hamiltonian of itinerant Majorana fermions in a given layer, $H^{l}$, is given by the first line of Eq.~(\ref{eq:Ham_MF}). The second line of Eq.~(\ref{eq:Ham_MF}) is instead the Hamiltonian of localized Majorana particles. As we show in~\cite{Note1}, they 
remain localized
for all values of parameters studied in the remainder of this paper. Thus, we will neglect 
them
in what follows. Direct calculation of $u^{l}_{\alpha}(\bm{r}_{i})$ is impractical, except in the simplest commensurate cases where translational invariance may be utilized to take them to be independent of $\bm{r}_{i}$, thus dramatically reducing the number of independent mean-field components. Fig.~\ref{fig:two} shows mean-fields $u^{t}_{\alpha,S}$ and $u^{b}_{\alpha,S}$ calculated as in Ref.~\cite{Seifert2018} for two stacking configurations, $S={\rm AA}$ and $S={\rm AB}$. For AB stacking, the boundaries between different phases appears shifted to values of $J/K$ which are roughly twice those found for the AA configuration. This finding can be interpreted as the result of the halved number of interlayer nearest-neighbours in AB bilayers.

In each commensurate configuration, the signs of mean fields can be flipped via a gauge transformation, thus only their moduli have physical meaning. Conversely, in a moir\`e superlattice one needs to fix these signs in such a way that, in certain limiting conditions, the behavior of the structure recovers the expected one. We set $u^{t,b}_{\alpha,{\rm BA}} = u^{t,b}_{\alpha,{\rm AB}}$, {\it i.e.} mean-fields have the same sign in AB and BA regions. This is dictated by the requirement that, in the limit of zero interlayer coupling, the Hamiltonian describes two independent Kitaev layers. 
We construct the twisted stack mean-fields by interpolating those of commensurate bilayers as $u^{l}_{\alpha}(\bm{r}_{i}) = \sum_{S} u_{\alpha, S}^{l} T(\bm{r}_{i}-\bm{r}_{S})$, where $\bm{r}_{S}$ denote positions of the three stacking configurations in the moir\`e unit cell [Fig.~\ref{fig:one}(a)]. We choose the periodic function $T(\bm{r}) = (1/9) \sum_{n = 1}^{3} \bigl( 1 + e^{i\bm{G}_{n} \cdot \bm{r}} + e^{-i\bm{G}_{n} \cdot \bm{r}} \bigr)$, to preserve the threefold rotational symmetry of the superlattice while matching the periodicity of the supercell. $\bm{G}_{1, 2}$ are the previously defined reciprocal lattice vectors for the moir\`e superlattice, and $\bm{G}_{3} = \bm{G}_{2} - \bm{G}_{1}$, while $T(\bm{r}) = 1$ [$T(\bm{r})=0$] for ${\bm r} = \bm{r}_{\rm AA}$ [${\bm r} = \bm{r}_{\rm AB},\bm{r}_{\rm BA}$]. Therefore, $u^{l}_{\alpha}(\bm{r})$ represents the simplest real-valued periodic function which smoothly interpolates between the values of 
mean-field parameters 
of AA, AB, and BA
regions.
The intralayer term can be written in momentum space representation as
\begin{equation}
\label{eq:Layer_K}
    \begin{split}
    H^{l} &= - \frac{i}{2} \sum_{\bm{k}} f^{l}(\bm{k}, \bm{0}) c^{\dagger}_{\bm{k}, l, A}c_{\bm{k}, l, B} 
    \\
    &- \frac{i}{6} \sum_{\bm{k},{\bm G}} f^{l}(\bm{k}, -\bm{G}) c^{\dagger}_{\bm{k}, l, A}c_{\bm{k}-\bm{G}_n, l, B},
    \end{split}
\end{equation}
where $\bm{k}$ is measured from the BZ center, ${\bm G}$ runs over the set $\pm {\bm G}_n$ ($n=1, 2,3$) and $f^{l}(\bm{k}, \bm{G}) = (1/3) \sum_{S} \sum_{\alpha} K^{\alpha} u^{l}_{\alpha, S}e^{i\bm{G}\cdot\bm{r}_{S}}e^{i(\bm{k} + \bm{G})\cdot\bm{\tau}^{l}_{\alpha}}$. We assume the following conventional~\cite{Bistrizer2011} 
form for interlayer interactions between itinerant Majorana fermions
\begin{equation}\label{eq:BM}
    H^{\perp} 
    = -iJ \sum^{2}_{n=0} \sum_{\bm{k},\mu\nu} w_{\alpha, \mu\nu}e^{i(\bm{g}_n^{t}\cdot\bm{\tau}^{t}_{\mu} - \bm{g}_{n}^{b}\cdot\bm{\tau}^{b}_{\nu})} 
    c^{\dagger}_{\bm{k}, t, \mu}c_{\bm{k}  - {\tilde {\bm G}}_n, b, \nu},
\end{equation}
where ${\tilde {\bm G}}_n \equiv \bm{G}_{n} - \Delta\bm{K}$, $\bm{g}_0 = \bm{G}_{0} = \bm{0}$, and $\Delta\bm{K} = (\bm{G}_{1} + \bm{G}_2)/3$. $w_{\alpha, {\rm AA}}$, $w_{\alpha, {\rm AB}}$, and $w_{\alpha, {\rm BA}}$ are parameters 
derived from the Majorana mean-field theories of
commensurately stacked bilayers, and $J$ is the Heisenberg exchange. Analogous to intralayer mean fields, we have $w^{l}_{\rm AB} = w^{l}_{\rm BA}$. We also take $w_{\alpha, {\rm BB}} = w_{\alpha, {\rm AA}}$. All terms simplify considerably in the isotropic limit where $K_{x} = K_{y} = K_{z} = K$: the mean-field components $u^{l}_{\alpha, S} = u^{l}_{S}$, and $w_{\alpha, S} = w_{S}$ become independent of bond direction. We note that although these mean-fields enter as free parameters in our equations, their values are determined by the (physical) couplings $J$ and $K$ via the solutions of the commensurate bilayers.

{\it Results}---The $\bm{k}$-dependent Hamiltonian is expanded in a plane-wave basis~\cite{Bistrizer2011, Koshino2018} and diagonalised to yield the itinerant Majorana fermion spectrum. We calculate the spectrum across the first MBZ around $(\bm{K}^{t}_{1}+\bm{K}^{b}_{1})/2$, and study the isotropic limit for simplicity. As seen in Fig.~\ref{fig:two}(a) the mean-fields (and therefore hopping amplitudes) depend on the ratio $J/K$, and as such the band structure is sensitive to changes in $J$. 
\begin{figure}
    \centering
\begin{overpic}[width=1\columnwidth]{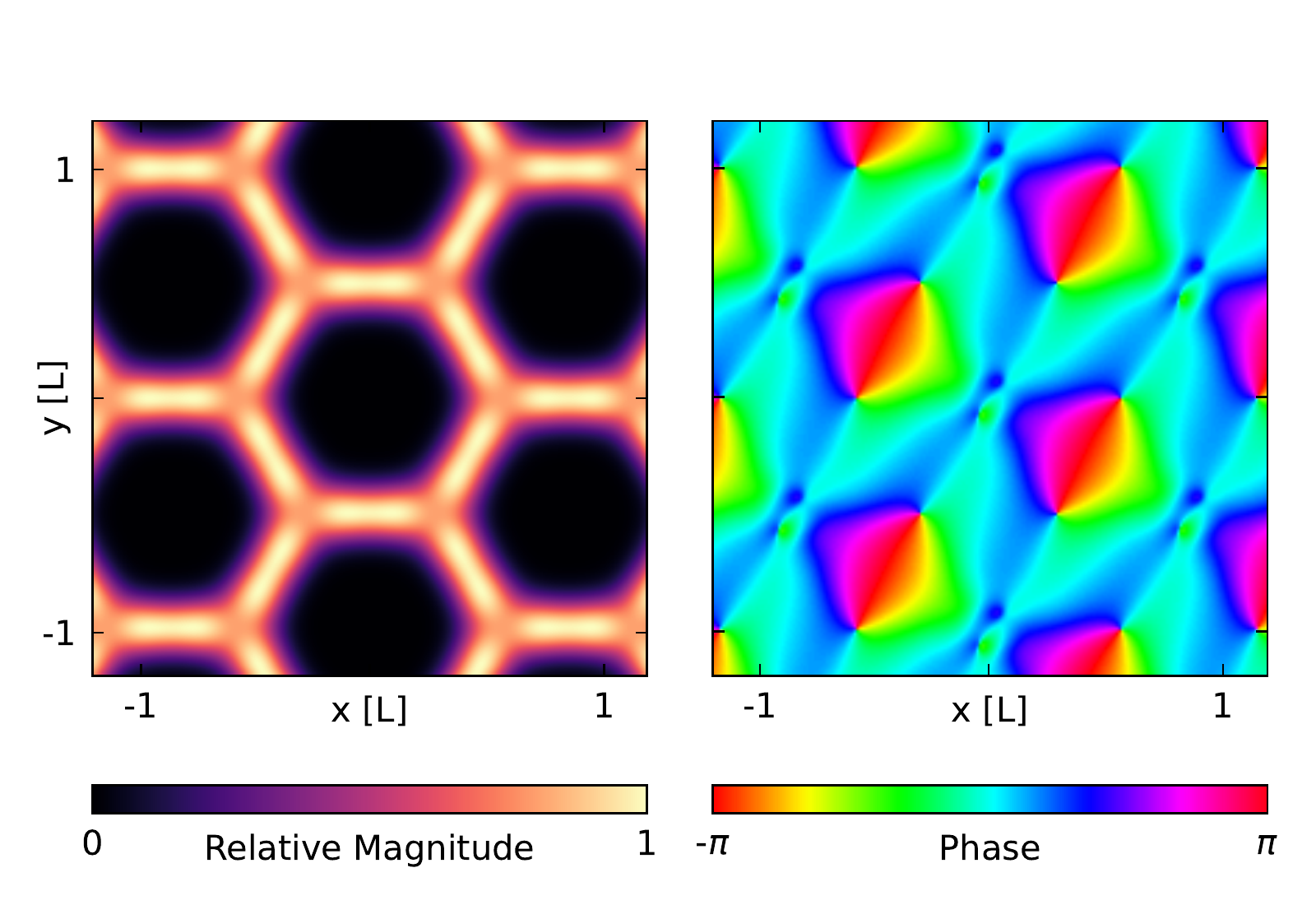}
\put(2,154){(a)}
\put(118,154){(b)}
\end{overpic}
    \vspace{-8mm}
    \caption{Panel (a) The local density of states (LDOS) around the energy of the first flat band ($E = 0.36K$).  Panel (b) Phase of the first flat band wavefunction at $\Gamma$, projected onto the A sublattice of the top layer. Here $J = 0.46K$. $L$ is the length of the moir\`e unit vector.}
    \label{fig:three}
\end{figure}
As shown in Fig.~\ref{fig:two}(a), for $J/K \lesssim 0.44$, both $w_{\rm AA}$, $w_{\rm AB} = 0$, and the system consists of two effectively decoupled monolayers with maximal intralayer 
mean-fields
$u^{t, b}_{S} = \pm 0.5$. We refer to the values of the mean-fields in this regime as the ``QSL'' values. We stress that the mean-field solution is exact in this case. The band structure obtained in this regime is shown in light gray in Figs.~\ref{fig:two}(b) and~(c) for reference. It corresponds to the folding of Dirac cones of isolated layers within the mini-Brillouin zone of a twisted (decoupled) bilayer. The band structure presents a linear low-energy dispersion around the corners of this smaller zone ($K$ and $K'$), and it is degenerate at higher energies at both the $M$ (middle point of the hexagonal mini-Brillouin zone) and $\Gamma$ points. In particular, higher-energy bands are six-fold degenerate at the latter.

For $0.44 \lesssim J/K \lesssim 0.58$, $u^{t, b}_{\rm AA}$ and $w_{\rm AA}$ deviate from their QSL values, while the mean-fields for the AB stacking remain unchanged, {\it i.e.} the interlayer couplings $w_{\rm AB} = w_{\rm BA}$ vanish exactly. Under these conditions the interlayer coupling becomes finite over the entire supercell, but is reduced in the neighborhoods of $\bm{r}_{\rm AB}$ and ${\bm r}_{\rm BA}$, and vanishes exactly at those points. Conversely, the intralayer hopping is minimum around AA regions. As $J/K$ enters in this range of values, the sixfold degeneracy at $\Gamma$ breaks, and bands split off and begin to flatten. There is no low-energy gap opening around the Dirac points even for large $J/K$, although the lowest band becomes increasingly energetically isolated from the rest.

Notably, very flat higher-energy bands appear close to other dispersive bands for $J/K \gtrsim 0.46$, thus forming a hyper-magic manifold~\cite{Scheer_arxiv_2022}. In the absence of modulation of intralayer hopping amplitudes, flat and dispersive bands are degenerate at the $\Gamma$ point. In Fig.~\ref{fig:two}, alongside the band structure, we also show the spinon density of states, which is seen to become very large at energies corresponding to flat bands. We expect these features to be reflected in, e.g., spin-dependent 2D-to-2D tunneling spectra~\cite{carrega_prb_2020}, which could thus be used to prove the existence of the hyper-magic manifold.

Visually, the group of bands above the lowest one in Fig.~\ref{fig:two}(c) strongly resembles the spectrum of a Kagome lattice. The latter tri-partite model exhibits two dispersive bands touching a perfectly-flat low-energy band at the $\Gamma$ point. To verify whether Kagome physics emerges in twisted Kitaev bilayers, we study the flat band wavefunction. Fig.~\ref{fig:three} shows the local density of states around the energy of the first flat band shown in Fig.~\ref{fig:two}(c). The wavefunction appears to be highly localised away from AA stacked regions, with maxima arranged in a trigonal coordination about each of the AB/BA stacked regions. Intriguingly, the localisation is akin to that of a Kagome lattice. It is also instructive to study the phase of the wavefunction. In Fig.~\ref{fig:three}(b) we plot the wavefunction of the first flat band at the $\Gamma$ point, projected onto the A sublattice of the top layer. We find phase vortex/anti-vortex pairs centered about the AB/BA stacked regions. This explains the apparent drop in the amplitude of the wavefunction at those points: certain components of the wavefunction must necessarily vanish there, due to the singular nature of the phase.

In~\cite{Note1} we show band structures calculated for higher values of $J/K$, which exhibit a higher degree of flattening. There, we also show results obtained for unphysical choices of mean-fields [{\it i.e.}, not constrained by the value of $J/K$ according to Fig.~\ref{fig:two}(a)]. We find that for some choices of intra- and inter-layer parameters, it is possible to make the flat band degenerate with the lowest one at the $\Gamma$ point. We also show band structures exhibiting a larger hyper-magic manifold similar to that seen in \cite{Scheer_arxiv_2022}, and demonstrate a mapping from our model to one of twisted-bilayer graphene equipped with complex interlayer couplings. These results highlight the higher degree of tunability of twisted multilayers of quantum-ordered materials compared to twisted graphene bilayers~\cite{He2021,Andrei2021}.

{\it Conclusion}---By considering variations in local stacking across a moir\`e heterostructure and their effect on mean-field parameters, we have constructed a mean-field approximation for a twisted bilayer of Kitaev quantum spin liquids. We have produced a working approximation for intra- and inter-layer tight-binding hopping amplitudes of deconfined Majorana particles in terms of solutions of mean-field theories on commensurately stacked bilayers. Accounting for these effects yields a complex band structure, which exhibits a hyper-magic manifold~\cite{Scheer_arxiv_2022} of flat bands and
Kagome-like
features 

Although at mean-field level relatively large values of $J/K$ are needed for the interlayer tunneling amplitude to become finite, in reality we expect the required $J$ to be much smaller. In fact, calculations beyond mean-field~\cite{Okamoto_prb_2013,Tomishige2018,Tomishige_prb_2019} show that the decoupled Kitaev phase is much less robust than predicted by mean-field theory. We stress that this ``fragility'' of the decoupled phase works in our favor, since it can make the hyper-magic manifold visible at smaller $J/K$. This in turn implies that the twisted Kitaev physics we describe can be observable in realistic van-der-Waals materials, where the interlayer Heisenberg coupling is not expected to be large. Note that the value of $J/K$ at which the hyper-magic manifold forms, can be tuned with the twist angle. This greatly improves the potential to experimentally realize the band flattening described here.

While preparing this manuscript, we became aware of a related work~\cite{Nica2022}.  While our results for small values of $J/K$ agree with Ref.~\cite{Nica2022}, we capture the effect of the varying stacking arrangement on the intralayer hopping, which was neglected in other works. We show that such terms can allow one to isolate flat from dispersive bands.

\begin{acknowledgments}
We acknowledge support from the European Commission under the EU Horizon 2020 MSCA-RISE-2019 programme (project 873028 HYDROTRONICS) and of the Leverhulme Trust under the grant RPG-2019-363. This work was also supported by the Engineering and Physical Sciences Research Council and the Graphene NOWNANO CDT.
\end{acknowledgments}

\appendix

\bibliography{refs}
\end{document}